\documentclass{camera}
\usepackage{graphicx}  

\begin{document}

%
\title{Ultra high energy cosmic rays: A review}

%
\author{Todor Stanev}

%
\organization{Bartol Research Institute, Department of physics and Astronomy,
 University of Delaware, Newark, DE 19716, U.S.A.}

\maketitle

\begin{abstract}
  We present the main results on the energy spectrum and composition
 of the highest energy cosmic rays of energy exceeding 10$^{18}$ eV
 obtained by the High Resolution Fly's Eye and the Southern
 Auger Observatory. The current results are somewhat contradictory
 and raise interesting questions about the origin and character
 of these particles. 
\end{abstract}

%

\section{Introduction}
  There is not an exact definition which cosmic rays should
 be called of ultrahigh energy. Generally the term is applied 
 to those cosmic rays that we think are not accelerated in our
 Galaxy, i.e. are of extragalactic origin. This definition is
 certainly model dependent, but it always includes cosmic
 rays of energy above 10$^{19}$ eV. Showers generated by  such
 high energy cosmic rays cosmic rays were first
 detected by the Vulcano Ranch air shower array in New Mexico,
 U.S.A.~\cite{JL1}. The first cosmic ray of energy 10$^{20}$ eV,
 and possibly higher, was also detected by the same array~\cite{JL2}.
 In this article we will call all particles of energy above 10$^{18}$ eV
 ultrahigh energy cosmic rays, abbreviated to UHECR. 

  At the time (1963) this did not cause any surprize: all physicists
 expected that the cosmic ray spectrum will go forever to higher
 and higher energies and the only reason we do not see such 
 particles is that their flux is very low. It was only after the
 discovery of the microwave background that Greisen~\cite{Greisen}
 in U.S. and Zatsepin \& Kuzmin~\cite{ZK} in the Soviet Union 
 predicted the end of the cosmi ray spectrum because of the
 cosmic rays interactions with the microwave background. This 
 feature is now called the {\em GZK} feature or cutoff.

\begin{figure}[thb]
\centerline{\includegraphics[width=10truecm]{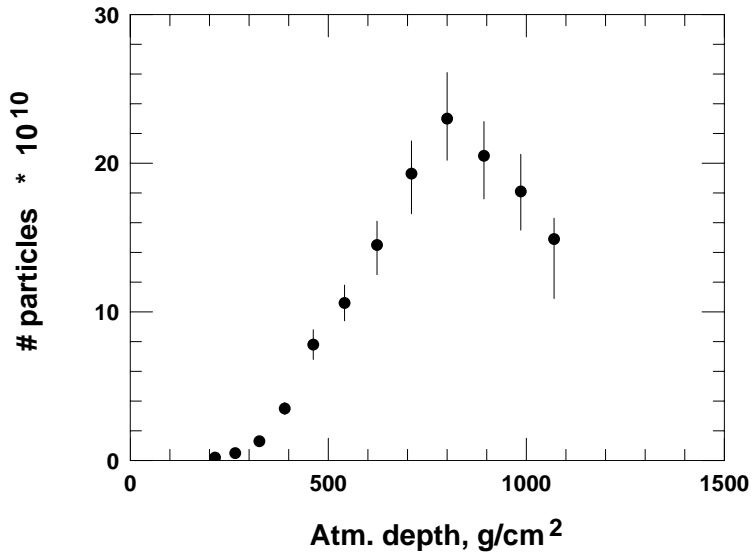}}
\caption{ Shower profile of the highest energy shower
 (3$\times$10$^{20}$ eV) detected by the Fly's Eye experiment.}
\label{highest} 
\end{figure}

  In order to give the reader an impression of how big these
 showers are we show in Fig.~\ref{highest} the shower
 profile (number of particles as a function of the atmospheric
 depth) of the highest energy cosmic ray shower detected
 by the Fly's Eye experiment~\cite{highest}. The energy estimate
 for this shower is 3$\times$10$^{20}$ eV and the number
 of shower electrons at the shower maximum development ($N_{max}$)
 is more than 2$\times$10$^{11}$. The energy estimate of such showers
 consistes of an integration on the shower profile times the average
 electron eneregy.

\begin{figure}[thb]
\centerline{\includegraphics[width=10truecm]{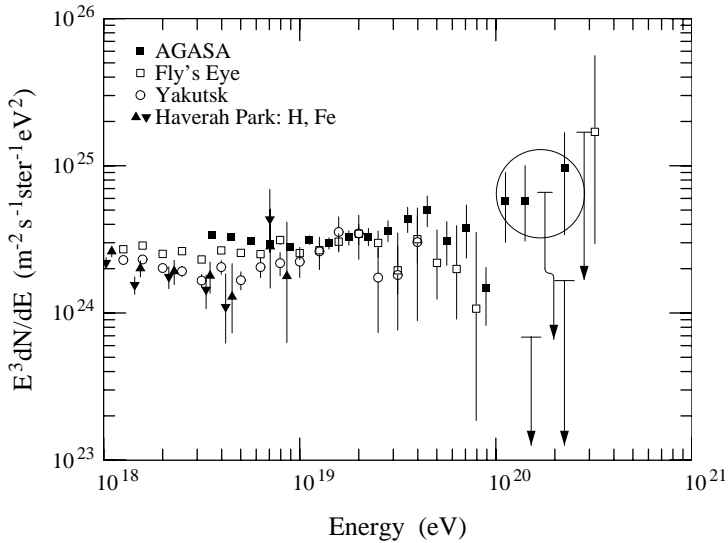}}
\caption{ UHECR spectrum as determined before the current generation
 of experiments, HiRes and Auger.}
\label{old_spectrum} 
\end{figure}

 The next figure shows the state of our knowledge of the energy
 spectrum of the UHECR before the
 current generation of experiments, HiRes and Auger. One of the data
 sets shown in Fig.~\ref{old_spectrum}, Agasa, attracts the attention
 since it does include more than 10 events of energy above 10$^{20}$
 eV~\cite{Agasa}.
 At the time Agasa was the biggest (100 sq.~km.) air shower array.
 Since we do not expect UHECR of this energy to be able to propagate
 from their sources to us, this result called for UHECR production
 models different from acceleration at astrophysical objects.
 As a result tens of exotic models explaining the existence of these
 particles as a result of ultraheavy $X$-particles decay appeared in 
 the literature. The UHECRs in shuch models are most likely to be
 either $\gamma$-rays or neutrinos because most of the decay 
 products have to be mesons rather than nucleons.

  Whatever the origin of the UHECR, acceleration at astrophysical objects
 or $X$-particle decay it is important to note that the energy spectrum
 detected at Earth is not the same as the acceleration and production
 spectrum is a result of the UHECR energy loss in propagation from their
 sources to us. We can use the detected energy spectrum and cosmic
 rays propagation calculations to study what the production spectrum
 is. Although all interaction properties are well known, 
 propagation calculations still have to make simplifying assumptions:\\ 
 \ \ \ \ $\bullet$ All sources have identical acceleration/production 
 spectra, and\\
 \ \ \ \ $\bullet$ Sources are isotropically distributed in the Universe.

\section{UHECR energy spectrum}

 In this section we will discuss the comtemporary measurements of
 the UHECR energy spectrum that were done by the High Resolution
 Fly's Eye (HiRes) detector and the Southern Auger Onservatory.
 The HiRes experiment consists of two fluorescent telescopes,
 HiRes 1 and 2. Fluorescent light is emitted by the air Nitrogen
 atoms excited by the ionization of the shower particles. The ligh
 emission is isotropic and the yield at see level is 4 UV photons
 per one meter of electron track. This yeld depends on the
 air density and temperature but 4 photons per meter of track is
 good enough for a simple idea for the photon fluxes at the telescope.
 HiRes~1 looks at elevations between 3 and 17$^o$
 above the horizon, while HiRes~2 doubles the field of view up
 to elevation of 31$^o$. The telescopes work both individually 
 or in stereo mode. Stereoscopic detection makes the shower
 analysis much more accurate. The energy estimate couples the
 integral of the shower profile and the average electron energy.

 The Southern Auger Observatory is a hybrid detector cosisting 
 of a huge air shower array of total area 3,000 km$^2$ and 24
 fluorescent telescopes that observe the air above it.
 The detectors of the surface array (SD) are water Cherenkov 
 tanks that register the Cherenkov light of the shower particles
 that hit the tank. Each water Cherenkov tank has a surface
 area of 10 m$^2$ and a depth of 1.2 m and is viewed by 3
 photomultiplier tubes.
 The  fluorescent telescopes are organized in 4 stations that
 occasionally can observe the fluorescent light in stereo. The
 surface array is fully efficient above shower energy of
 3$\times$10$^{18}$ eV.
 The energy estimate of the surface array is obtained by
 the correlation of the shower signal at 1000 m from the shower
 axis, $S_{1000}$ to the fluorescent signals in events detected
 by both detectors. The lower energy part of the spectrum is measured 
 by hybrid events, where at least one of the surface detectors 
 triggered in coincidence with the fluorescent telescope.
 
\begin{figure}[thb]
\centerline{\includegraphics[width=10truecm]{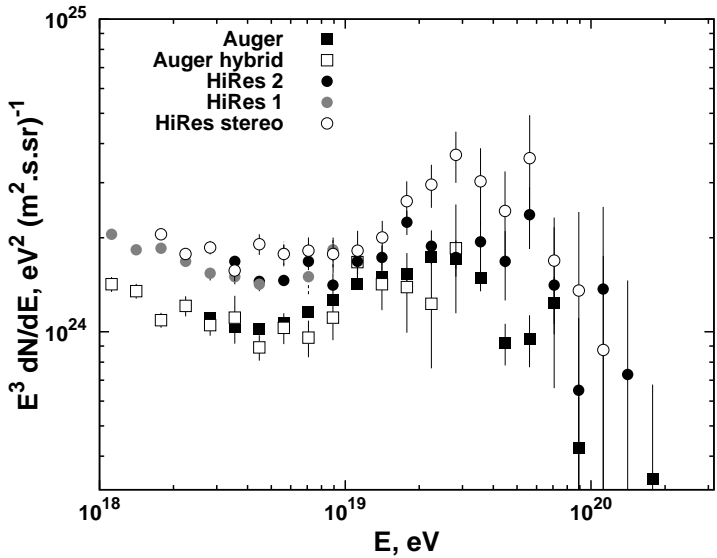}}
\caption{ UHECR spectrum as measured by HiRes and Auger.}
\label{spectrum} 
\end{figure}

 Fig.~\ref{spectrum} shows the cosmic ray spectrum as measured
 by HiRes and Auger. The independ analyses of HiRes~1 and
 2~\cite{HiRes_prl} are shown with black and grey points while
 the stereo analysis~\cite{HiRes_stereo} is shown with
 empty circles. The Auger surface array spectrum~\cite{Auger_prl}
 is shown with black squares and the hybrid measurement~\cite{Auger_hybrid}
 is shown with empty squares.
 Since the surface detector has much higher statistics the
 hybrid data are only important for energies below 10$^{19}$ eV.

 The first conclusion from Fig.~\ref{spectrum} is that both
 experiments confirm the GZK cutoff. After energy of about
 4$\times$10$^{19}$ eV the cosmic ray spectrums steepens and
 there are very few events above that energy. Note that 
 the measured spectrum is multiplied by E$^3$ which makes 
 the small differences in the energy calibration look 
 significantly bigger. There are, though, some differences
 that steer the interpretations of the energy spectra towards 
 different models. The Auger energy spectrum is fitted to 
 E$^{-2.6}$ behaviour over an order of magnitude above
 3$\times$10$^{18}$ eV, while the HiRes spectrum is somewhat 
 steeper. The same is true for the exact position of the GZK
 cutoff, where the experimental statistics is low.

  In spite of the similarity of the two spectra they allow quite 
 different interpretations for the origine and type of UHECR.
 The HiRes spectrum is fully consistent with the model of
 Berezinsky et al~\cite{beretal05} which assumes a steep cosmic ray
 acceleration spectrum (E$^{-2.7}$) and a pure proton composition.
 There is no need for a cosmological evolution of the cosmic
 ray sources.

  The interpretation of the Auger energy spectrum is much more 
 complicated as it allows several different models. The first one
 is not dissimilar to that of Berezinsky et al - pure proton 
 composition, E$^{-2.55}$ acceleration spectrum and no cosmological
 distribution of the UHECR sources. The second proton composition
 model  requires flatter E$^{-2.3}$ acceleration spectrum, a very strong
 (proportional to $(1 + z)^5$) cosmological evolution of the sources.
 The third model is that of mixed composition, i.e. the same nuclei
 that exist in the galactic cosmic rays are accelerated at the 
 powerful extragalactic sources. The acceleration spectra are relatively
 flat and the exact parameters depend on the composition of the
 accelerated cosmic rays.

\section{Cosmic ray composition}

 The measurement of the cosmic ray composition at high energy
 depends on the parameters of the showers that different nuclei 
 generate.  At energy around 10$^{15}$-10$^{16}$ eV the main
 composition parameter is the ratio of muons to electrons in
 the shower. At higher energy such a measurement becomes
 difficult since the muon counters are much more expensive
 and the main composition measurement is the depth of the shower
 maximum development X$_{max}$. Using a very simple analytic
 shower model Matthews~\cite{Matthews2005} showed that
 $$ X^A_{max}\; = \; X^p_{max}\; - ; X_0 ln A$$ where $X_0$ is
 the radiation length.
 Showers induced by heavy nuclei develop significantly 
 faster than those of protons. The contemporary fluorescent 
 detectors can measure X$_{max}$ with an accuracy of about 20 g/cm$^2$
 while the difference between the average  X$_{max}$ in proton 
 and iron showers is about 100 g/cm$^2$.

  Showers initiated by $\gamma$-rays or neutrinos have also 
 development characteristics that are different from those of
 showers initiated by nuclei. Both HiRes~\cite{HR_neu} and
 Auger~\cite{Auger_neu} were able to put limits on the fluxes
 of ultrahigh energy neutrinos. These limits were set on specific
 neutrino flavors but the limit the total neutrino flux since
 oscillations would led to $\nu_e$ : $\nu_\mu$ : $\nu_\tau$ ratio
 of approximately 1 : 1 : 1.

  The Auger Observatory also put limits on the fraction of $\gamma$-rays
 in the total cosmic ray flux. The limits come from the fact that
 very high energy $\gamma$-ray showers develop significantly deeper 
 in the atmosphere than proton showers do. Two different techniques 
 were used again: hybrid showers between 10$^{18}$ and 10$^{19}$
 eV~\cite{Auger_g1}
 and surface detector~\cite{Auger_g2} above that energy.
 The fraction of $\gamma$-rays
 in the cosmic ray flux above 2$\times$10$^{18}$ eV was limited to
 less than 4\% and the fraction above 10$^{19}$ eV to 2\%.
 These limits almost eliminate the exotic models for UHECR production
 although it is still possible that highest energy particles, where
 the statistics is too low to set limits, are still $\gamma$-rays.

\begin{figure}[thb]
\centerline{\includegraphics[width=10truecm]{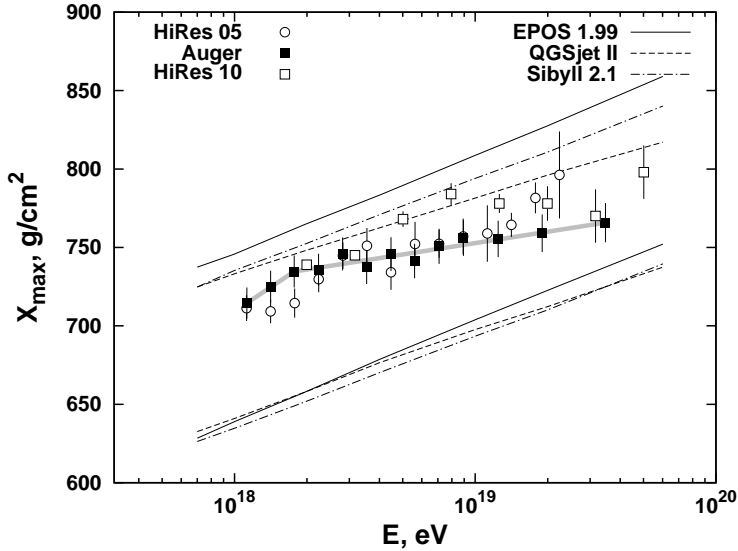}}
\caption{Depth of maximum as a function of the shower
 energy as measured by HiRes and Auger and the predictions 
 of three different interaction models for proton and iron
 showers.}
\label{compos} 
\end{figure}

  An important parameter in the study of the shower depth of maximum
 is the elongation rate $D_{10}$ - the rate of change of X$_{max}$ per
 a decade of energy. $D_{10}$ is approximately (1 - B)X$_0$ln10, where
 B is a parameter that depends on the hadronic interaction model.
 For different models D$_{10}$ is between 50 and 60 g/cm$^2$ if the
 cosmic ray composition is constant. In case that the composition
 becomes lighter with energy D$_{10}$ grows. If it becomes heavier
 D$_{10}$ is lower. Fig.~\ref{compos} shows the results on X$_{max}$
 energy dependence as measured by the HiRes and Auger together with the
 predictions of three different interaction models for proton and Fe
 showers.     
  
 The points labeled HiRes 05 show the first result with relatively 
 small statistics published by HiRes in 2005~\cite{HR2005}. These
 points imply that the shower depth of maximum increases faster 
 than any of the interaction models predict. This suggests that
 the cosmic ray composition becomes lighter with energy, although
 it can be heavier than pure protons, especially if the EPOS 1.99
 model is correct. The black squares show the X$_{max}$ results
 of the Auger Observatory~\cite{Auger_comp}. The general behavior is
 shown with the gray line. The three lowest energy points suggest 
 D$_{10}$ of 106 g/cm$^2$ with a large error bar. At higher energy
 the elongation rate becomes 24$\pm$3 g/cm$^2$ . The interpretation
 in terms of cosmic ray composition should be that up to
 3$\times$10$^{18}$ eV the cosmic ray composition becomes lighter
 and at higher energy it becomes heavier. The highest energy point
 at about 4$\times$10$^{19}$ eV is closer to iron than it is to
 protons in any of the interaction models. The open squares 
 show the HiRes data set taken in stereo~\cite{HR10}. Some of these points
 are even higher than expected for Fe showers in the QGSJet II
 model. There is an obvious disagreement between the two experiments.

\begin{figure}[thb]
\centerline{\includegraphics[width=10truecm]{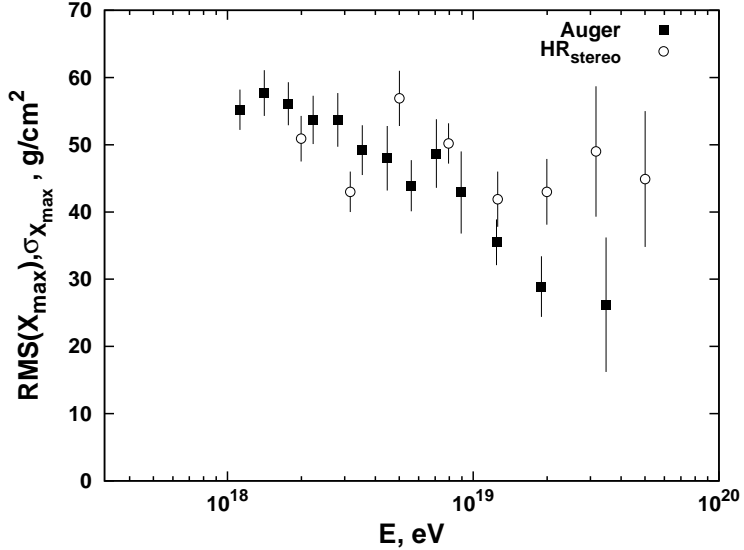}}
\caption{Width of the X$_{max}$ distributions measured by Auger and 
 HiRes.}
\label{RMS} 
\end{figure}

 This disagreement carries over when the two experiments examined
 the width of the X$_{max}$ distributions in each of the energy 
 bins shown in Fig.~\ref{RMS}. The width of these distributions also
 reflects the cosmoc ray composition. The predictions for proton
 showers are of order 60 g/cm$^2$ while for Fe showers they are about
 20 g/cm$^2$. In the case of Auger the decrease of the rms values 
 follow the average depth of maximum and both of them suggest
 a composition that becomes heavier with the increasing energy,
 In the case of HiRes the best fit is a straight line. Note that
 the definitions of {\em width} are not identical. Auger uses 
 directly the rms value of the distribution while HiRes gives the
 Gaussian width after the long tale of the distribution is cut off. 

 One possibility for the disagreement is that HiRes and Auger select
 their event samples in different ways. The reason is that none of
 the fluorescent detectors observes the whole sky and only elevations
 up to 31$^o$ in the best case. The experiments want to be certain
 they do are not biased toward early or late developing showers and
 apply very different cuts to the data. Auger, for example only uses
 hybrid events which implies selecting showers relatively close to
 the fluorescent
 detector while HiRes does not use showers closer than 10 km to any
 of their telescopes. 
 
\section{Arrival directions of the highest energy cosmic rays}

 The idea is that UHECR do not scatter much in the galactic and
 extragalactic magnetic fields so they will cluster arround their
 sources and thus reveal them. The Southern Auger Observatory 
 is the biggest UHECR detector and has the highest chance to look
 for the sources of these particles. Early in the game Auger 
 made a trial scan, identified the procedure to follow and in 
 2007 they published a paper~\cite{Auger_science} on the 
 correlation of their events of energy exceeding 57 EeV with the
 active galactic nuclei from the VCV~\cite{VCV} catalog at redshifts
 smaller than 0.018.

\begin{figure}[thb]
\centerline{\includegraphics[width=12.5truecm]{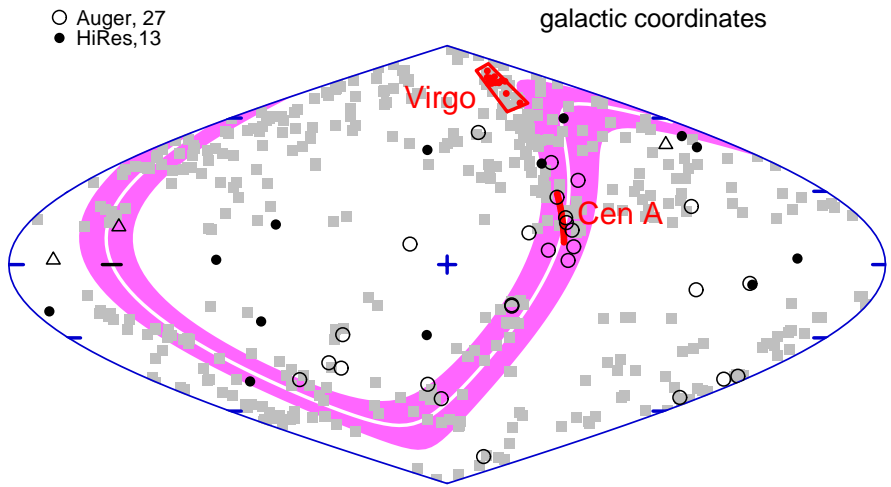}}
\caption{Correlation of the arrival directions of the Auger
 and HiRes highest energy events with the VCV catalog.}
\label{corr} 
\end{figure}

 Out of 27 highest energy events 19 events came from directions
 not exceeding 3.1$^o$ from an AGN. Most of the events that did
 not correlate passed less than 12$^o$ from the galactic plane
 where the galactic magnetic field is the strongest. Accounting
 for the scans in particle energy, AGN distance, and distance 
 from the AGNs the significance of the correlation was not huge,
 but still exceeding 3$\sigma$. These events and the AGNs 
 with $|b|$ > 12$^o$ are shown in Fig.~\ref{corr}.
 
 HiRes looked for correlation of their highest energy 13 events
 with the AGNs from the same catalog and did not see any
 correlation~\cite{HiRes_corr}.
 The fields of view of the two experiments do not coincide, but
 there is enough overlap to look for correlations of both data sets.
 
 The Auger Observatory has more than doubled its statistics by the
 time of of International Cosmic Ray Conference in 2009. The level of
 correlation, however, decreased significantly to about
 2$\sigma$~\cite{Auger_corr09}.
 Since the VCV catalog does not appear to be the best one to contain
 possible UHECR sources the collaboration attempted to correlate the
 arrival directions of their highest energy events with objects
 from different ones with similar results. Only about 30\% of 
 UHECR seem to correlate with the directions of powerful extragalactic
 objects while the rest seem to come from an isotropic distribution.
 The exception is the direction of the nearby (less than 4 Mpc)
 radio galaxy Centaurus A around which several events are clustered.

 There is obviously an internal contradiction between the idea of 
 correlation of the arrival directions with extragalactic objects
 and that of heavy cosmic ray composition. If the highest energy events 
 are indeed heavy nuclei they may scatter a lot and appear to come
 to us from isotropic directions. The question of UHECR astronomy
 can most likely be solved only with vastly increased statistics.
 Auger is now proposing a new, much larger Northern Observatory.

\section{Summary}
 
 The main question raised by the results of the Agasa collaboration,
 the energy extent of the cosmic ray spectrum, is now solved.
 Both HiRes and Auger observe its end, the GZK cutoff. With
 a small change in the energy assignment, less than the systematic
 error of 20\%, the two measured spectra will be fully consistent.

 The results on the cosmic ray composition are much more 
 contradictory. HiRes sees cosmic ray composition close to 
 a purely proton one while Auger observes increasily heavier
 composition above 3$\times$10$^{18}$ eV. The reasons for 
 this disagreement are not obvious and it will take the two 
 groups lots of work and collaboration to understand them.
 There maybe some help from the new hybrid experiment
 Telescope Array which combines scintillator counters with
 fluorescent telescopes.

 The question about the sources of these particles is not yet
 solved. Less than 1/3 of the highest Auger events correlate
 in arrival direction with the possible sources from different
 catalogs. The solution will most likely require significant
 increase of the experimental statistics.\\[3truemm]
{\bf DISCUSSION}\\[3truemm]
{\bf DANIELE FARGION} Why we do not see UHECR events in Auger
 towards Virgo and why are no events toward Norma?\\
{\bf TODOR STANEV} The answer of Auger is that after accounting 
 for the exposure (1/3) and for the dostance (1/25) they expect
 75 times less events from Virgo for equal luminosity. There are
 models, including yours, that can explain the lack of events.\\[3truemm]
{\bf SERGIO PETRERA} I have a comment on the HiRes spectrum -
 They show that their data are well fit by the model of
 Berezinsky et al. They never published fits with other models, but 
 mixed composition models can successfully fit data as well because of the
 many handles they have.\\
{\bf TODOR STANEV} The HiRes group used an existing model to fit their
 spectrum. After it worked they did not present other fits.\\[3truemm]
{\bf FRANCESCO VISSANI} Greisen, Zatsepin and Kuzmin put together their
 proposal {\em before} the data were known. Many of the models you
 discussed were proposed {\em after} the data were available, for instance
 the superheavy decaying particle model. Don't you believe that this
 consideration already puts these theories in a different position?\\
{\bf TODOR STANEV} You are correct. The GZK was the original model.
 Contemporary models are made to fit observations. It would be better
 to call them fits to data rather than models.


\end{document}